\documentclass[preprint]{sigplanconf}
\usepackage{mdwtab}
\usepackage{multirow}
\usepackage{amsmath}
\usepackage{amssymb}
\usepackage{trfrac}
\usepackage{color}
\usepackage{parskip}
\usepackage{tikz}
\usetikzlibrary{matrix,arrows}
\usepackage{microtype}               
\usepackage{amsthm}
\theoremstyle{marginbreak}
\newtheoremstyle{megacz}
  {\parskip}
  {0pt}
  {}
  {0pt}
  {\bfseries}
  {}
  {\parskip}
  {}
\renewenvironment{proof}[1][\proofname]{

{\it#1.} }{\hfill $\Box$}
\theoremstyle{megacz}
\newtheorem{theorem}{Theorem}
\newtheorem{remark}{Remark}
\newtheorem{lemma}{Lemma}

\newtheorem{definition}{Definition}

\usepackage{verbatim}
\usepackage{fancyvrb}
\def\tild{\rlap{\lower.85ex\hbox{\large\char126}}\ }
\DefineVerbatimEnvironment%
 {verbatim}%
 {Verbatim}%
 {codes={\catcode`~=\active},%
  samepage=true,
  defineactive=\def~{\tild}%
 }

\renewcommand{\topfraction}{0.9}
\renewcommand{\floatpagefraction}{0.7}

\def\subst#1#2{{[#1{\ttt{:=}}#2]}^G}
\def\veta{{\vec\eta}}
\def\RN#1{\multirow{2}{*}{\textsf{#1}}}
\def\rospace{2.5mm}
\def\rospacex{1.81mm}
\def\interrow{\\ \vgap{\rospace}\hline\vgap{\rospace}}
\def\interrowx{\\ \vgap{\rospacex}\hline\vgap{\rospacex}}
\def\To{\Rightarrow}
\def\Uuc#1{#1}
\def\ttt#1{\text{\tt{#1}}}
\def\tsf#1{\text{\textsf{#1}}}
\def\code#1{\langle\hspace{-0.7mm}[{#1}]\hspace{-0.7mm}\rangle}
\def\G{\Gamma}
\def\C{{\mathbb C}}
\def\K{{\mathbb K}}
\def\M{{\mathbb M}}
\def\O{{\mathbb O}}
\def\D{{\mathbb D}}
\def\E{{\mathbb E}}
\def\curry{\text{{\textsf{curry}}}}
\def\eval{\text{{\textsf{eval}}}}
\def\lift{\text{{\textsf{lift}}}}
\def\id{\text{{\textsf{id}}}}
\def\strength{\text{{\textsf{strength}}}}
\def\first{\ttt{first}}
\def\second{\ttt{second}}

\def\>{\ttt{>>>}}
\def\ctree#1#2{#1,#2}
\def\defeq{\stackrel{\text{\tiny def}}{\equiv}}
\def\fo{\tsf{firstClass}}
\def\int{\ttt{int}}

\def\GArrow{{\tt GArrow}}
\def\Arrow{{\tt Arrow}}
\def\GArrows{{\tt GArrow}s}
\def\Arrows{{\tt Arrow}s}

\begin{document}

\conferenceinfo{WXYZ '10}{date, City.} 
\copyrightyear{2010} 
\copyrightdata{[to be supplied]} 
\titlebanner{}
\preprintfooter{}

\title{Multi-Stage Programs are Generalized Arrows}
\subtitle{
\textcolor{red}{
This paper is obsolete and has been superceded by\\
{\it Multi-Level Programs are Generalized Arrows} \\
available here:\\
{\tt http://arxiv.org/pdf/1007.2885}
}
}

\authorinfo{Adam Megacz}
           {UC Berkeley}
           {{\tt megacz@berkeley.edu}}

\maketitle

\begin{abstract}
The lambda calculus, subject to typing restrictions, provides a syntax
for the internal language of cartesian closed categories.  This paper
establishes a parallel result: staging annotations
\cite{TahaSheard:MetaML}, subject to named level restrictions, provide a
syntax for the internal language of Freyd categories, which are known
to be in bijective correspondence with \Arrows.  The connection is
made by interpreting multi-stage type systems as indexed functors from
polynomial categories to their reindexings
(Definitions~\ref{definition-reification}~and~\ref{contemplation}).

This result applies only to multi-stage languages which are (1)
homogeneous, (2) allow cross-stage persistence and (3) place no
restrictions on the use of structural rules in typing derivations.
Removing these restrictions and repeating the construction yields {\it
  generalized arrows}, of which \Arrows\ are a particular case.  A
translation from well-typed multi-stage programs to single-stage
\GArrow\ terms is provided.  The translation is defined by induction
on the structure of the proof that the multi-stage program is
well-typed, relying on information encoded in the proof's use of
structural rules (weakening, contraction, exchange, and context
associativity).

Metalanguage designers can now factor out the syntactic machinery of
metaprogramming by providing a single translation from staging syntax
into expressions of generalized arrow type.  Object language providers
need only implement the functions of the generalized arrow type class
in point-free style.  Object language users may write metaprograms
over these object languages in a point-ful style, using the same
binding, scoping, abstraction, and application mechanisms in both the
object language and metalanguage.

This paper's principal contributions are the {\tt GArrow} definition
of Figures~\ref{garrow-defs} and \ref{laws}, the
translation in Figure~\ref{core} and the category-theoretic semantics
of Definition~\ref{definition-reification}.  An accompanying Coq proof
formalizes the type system, translation procedure, and key theorems.
\end{abstract}


\begin{figure*}
\begin{minipage}[b]{.48\linewidth}
\begin{verbatim}
Class Arrow
            ((~>):Set->Set->Set) :=






 arr    : (a->b) -> (a~>b)

 (>>>)  : a~>b -> b~>c -> a~>c
 first  : a~>b -> (a*c)~>(b*c)

 (~~)   : a~>b -> a~>b -> Prop

 pf1    : Equivalence (a~~b)
 pf2    : Morphism (a~~b ==> b~~c ==> a~~c) (>>>)
 pf3    : Morphism (a~~b ==> (a*c)~~(b*c)) first
\end{verbatim}
\caption{Definition for the {\tt Arrow} class.  See also Remark~\ref{syntax-remark}.\label{arrow-defs}}
\end{minipage}
\begin{minipage}[b]{.48\linewidth}
\begin{verbatim}
Class GArrow ((**):Set->Set->Set)
             ((~>):Set->Set->Set) :=

 id     :         a ~> a
 assoc  : (a**b)**c ~> a**(b**c)
 cossa  : a**(b**c) ~> (a**b)**c
 copy   :         a ~> a**a
 drop   :      a**b ~> a
 swap   :      a**b ~> b**a

 (>>>)  : a~>b -> b~>c -> a~>c
 first  : a~>b -> (a**c)~>(b**c)

 (~~)   : a~>b -> a~>b -> Prop

 pf1    : Equivalence (a~~b)
 pf2    : Morphism (a~~b ==> b~~c ==> a~~c) (>>>)
 pf3    : Morphism (a~~b ==> (a**c)~~(b**c)) first
\end{verbatim}
\caption{Definition for the {\tt GArrow} class.  See also Remark~\ref{syntax-remark}.\label{garrow-defs}}
\end{minipage}
\end{figure*}

\section{Introduction}

Metaprogramming, the practice of writing programs which construct and
manipulate other programs, has a long history in the computing
literature.  However, prior to \cite{PfenningLee:LEAP} little of it
dealt with metaprogramming in a statically typed setting where one
wants to ensure not only that ``well typed programs do not go wrong,''
but also that well typed metaprograms {\it do not produce ill-typed
  object programs}.

One of the most popular applications of statically typed
metaprogramming has been the use of monads to account for different
{\it notions of computation} \cite{Moggi:NotionsOfComputation} as the
impure programs manipulated by pure functions in a category equipped
with a Kleisli triple.  The use of monads in functional programming
was later generalized to \Arrows\ by Hughes, who writes ``every time
we sequence two monadic computations, we have an opportunity to run
arbitrary code in between
them. \cite{Hughes:GeneralisingMonadsToArrows}'' \Arrows\ curtail this
freedom, permitting the inclusion of static information.  In practice,
this has made \Arrows\ a popular framework for metaprogramming,
particularly when one is allowed to do things with object programs
other than run them.

Because adding a new object language involves nothing more than
implementing the functions required by the \Arrow\ type class, this
approach to embedding makes it quite easy to {\it provide} new object
languages.  Although all embedded languages share a common syntax
\cite{Paterson:NewNotationForArrows}, this syntax is profoundly
different from that of the metalanguage, which can make it difficult
to {\it use} object languages.

By contrast, staging annotations \cite{TahaSheard:MetaML} embed an
object language within the metalanguage using the same binding,
scoping, abstraction, and application mechanisms as the metalanguage.
However, the type system of the metalanguage must reflect the type
system of the object language, so adding a new object language is
quite difficult and generally requires making modifications to the
metalanguage compiler.

This paper will use, as a running example, the {\tt pow} function
which has become ubiquitous in the metaprogramming literature.  Here
is the {\tt pow} program written using \Arrow\ notation
\cite{Paterson:NewNotationForArrows}:
\begin{verbatim}
pow n =
  if n==0
  then cst 1
  else proc x ->
       do pow'    <- (pow (n-1)) -< x
          result  <- (*)         -< (x, pow')
          returnA -< result
\end{verbatim}
Here is an equivalent program written using staging annotations:
\begin{verbatim}
pow n x =
  if n==0
  then <[ 1 ]>
  else <[ ~x * ~(pow (n-1) x) ]>
\end{verbatim}
Section~\ref{arrows-section} reviews \Arrows\ and introduces
generalized arrows.  Section~\ref{staging-section} presents a grammar
and type system for a simplified MetaML-style \cite{TahaSheard:MetaML}
multi-stage programming language.  Section~\ref{the-translation}
provides a translation procedure which produces generalized arrow
values from the typing derivations of well-typed multi-stage programs.
Section~\ref{examples-section} walks through a few example programs,
and Section~\ref{categorical-perspective} formalizes the
category-theoretic underpinnings of staging annotations.

\section{\Arrows}
\label{arrows-section}

From a programmer's perspective, an \Arrow\ is a type belonging to the
Coq type class \cite{SozeauOury:FirstClassTypeClasses} shown in
Figure~\ref{arrow-defs}.  Briefly, the members of the class are type
operators {\tt ({\tild}>)} which take two arguments, supplied along with
a function {\tt arr} which lifts arbitrary functions into \Arrows, a
function {\tt (>>>)} which composes \Arrows, and a function {\tt
  first} which lifts an \Arrow\ on a given type to an \Arrow\ on tuples with
that type as the first coordinate and the identity operation on the
second coordinate.  The last four declarations define an equivalence
relation {\tt (\tild\tild)} and require that {\tt (>>>)} and {\tt
  first} preserve it.

\begin{remark}
\label{syntax-remark}
To improve readability, the following elements of Coq syntax have been
elided from the printed version of this paper: semicolons, curly
braces, {\tt Notation} clauses, {\tt Implicit Argument} clauses,
explicit instantiation of implicit arguments, and polymorphic type
quantifiers (specifically, {\tt forall} occurring immediately after a
colon).  The complete Coq code, which includes the elided text, is
available online\footnote{{{\tt
      http://www.cs.berkeley.edu/{\tild}megacz/garrows/GArrow.v}}}
\end{remark}

\subsection{Generalized \Arrows\ (\GArrows)}

The Coq declaration for the {\tt GArrow} class is shown in
Figure~\ref{garrow-defs}; the laws for {\tt GArrows} can be found in
Figure~\ref{laws} using mathematical notation, and in
Figure~\ref{laws-coq} using Coq notation.  Proofs of these
propositions appear as obligations for any code attempting to create
an instance of the {\tt GArrow} class,  providing
machine-checked assurance that the laws are satisfied.

Comparing the two declarations, one can see that \GArrows\ generalize
\Arrows\ in two ways:
\begin{enumerate}
\item The {\tt arr} constructor is omitted, and part of its
      functionality is restored via {\tt id}, {\tt assoc}, {\tt
      cossa}, {\tt drop}, {\tt copy}, and {\tt swap}.

\item The methods of the \Arrow\ class are specified in terms of tuple
      types, which are assumed to be full cartesian products.
      \GArrows\ relax this restriction, assuming only that the tupling
      operator is a monoid.
\end{enumerate}
Parameterizing \GArrow\ over an arbitrary {\tt (**):Set->Set->Set}
operator rather than requiring the use of the cartesian product allows
for more generality: while there is a straightforward function of
type $(\forall\alpha)\alpha{\to}(\alpha,\alpha)$, there is no total
function of type
$(\forall\ttt{(**):Set->Set->Set})(\forall\alpha)\alpha{\to}(\alpha\ttt{**}\alpha)$.
The weaker construct makes it possible to deny users the ability to
form such functions where they are inappropriate.  In particular, it
prevents properties of the cartesian product from imposing
unwanted properties upon object language contexts, as will be shown in
Definition~\ref{definition-reification} and utilized in Section~\ref{biarrows}.

\begin{figure}
\begin{align}
\ttt{id} \ \>\  f  &=
   f
\\
f  \ \>\  \ttt{id} &=
   f
\\
(f\ \>\ g)\ \>\ h &=
   f\ \>\ (g\ \>\ h)
\\
\first\ (f\ \>\ g) &=
   (\ttt{first\ }f)\ \>\ (\ttt{first\ }g)
\\
\first\ (\first\ f)\ \>\ \ttt{assoc} &=
   \ttt{assoc}\ \>\ \first\ f
\\
\ttt{cossa} &=
  \ttt{swap}\ \>\ \ttt{assoc}\ \>\ \ttt{swap}
\\
\first\ f\ \>\ \ttt{drop} &=
   \ttt{drop}\ \>\ f
\label{first-drop-drop-f}
\\
\ttt{swap}\ \>\ \ttt{swap} &=
   \ttt{id}
\\
\ttt{copy}\ \>\ \ttt{swap} &=
   \ttt{copy}
\end{align}
\caption{ Generalized Arrow laws.  The first five laws are taken from
  \cite[Figure 1]{Paterson:NewNotationForArrows}.  The sixth law
  defines {\tt cossa} in terms of {\tt swap}; this makes it a
  redundant operation (much like {\tt ***} for {\tt Arrow}s), though
  Section~\ref{affine-linear-ordered} investigates variants which
  eschew {\tt swap}, making {\tt cossa} no longer redundant.  The
  seventh law expresses the fact that {\tt first} should not have side
  effects.  The last two laws establish some straightforward
  properties of {\tt swap} and {\tt copy}.  A Coq
  rendition of these laws can be found in Figure~\ref{laws-coq}.
\label{laws}}
\end{figure}

\begin{remark}
The following {\tt Arrow} laws from \cite[Figure
  1]{Paterson:NewNotationForArrows} have been omitted from {\tt
  GArrow} because they serve only to regulate {\tt arr}:
\begin{align}
\ttt{arr}(g\circ f) &=
  \ttt{arr}\ f\ \>\ \ttt{arr}\ g
\\
\first(\ttt{arr}\ f) &=
  \ttt{arr}(f\times\ttt{id})
\label{arr-f-f-id}
\\
\first\ f\ \>\ \ttt{arr}\ (\ttt{id}\times g) &=
  \ttt{arr}\ (\ttt{id}\times g)\ \>\ \first\ f
\end{align}
However, (\ref{arr-f-f-id}) above does serve the same purpose as law (\ref{first-drop-drop-f}) of Figure~\ref{laws}.
\end{remark}

\begin{theorem}
Every {\tt Arrow} is a {\tt GArrow prod}, where {\tt
  prod} is the cartesian product.
\begin{proof}
{\tt Instance Arrows\_are\_GArrows} in {\tt GArrow.v}
\end{proof}
\end{theorem}

\section{Staging Annotations}
\label{staging-section}

\subsection{Natural Deduction}

This section briefly reviews the structural rules for natural
deduction.  $\Delta$ will denote derivations, $\Sigma$ will denote
propositions and $\G$ will denote contexts, where a context consists
either of a single proposition or a pair of subcontexts:
$$
\G ::= \Sigma\ |\ \ctree{\G}{\G}
$$
Therefore contexts can be viewed as binary trees.

\begin{remark}
Although logically quite conventional -- the $(\ctree{\cdot}{\cdot})$
construct is exactly logical conjunction -- this choice is
proof-theoretically nonstandard; contexts are usually handled as
lists.  However, the translation given in
Section~\ref{the-translation} is only valid for proof derivations
which are completely explicit about every structural rule invocation.
The positions of these invocations in the proof derivation carry
information which is used by the translation.
\end{remark}

By representing contexts with binary trees rather than lists one can
avoid introducing rules which {\it implicitly} rearrange the context.  One example
of such a rule is one which uses ellipsis to abbreviate a sequence of
propositions: 
$$\G,\ldots,x:\tau\vdash\Sigma$$
Another example is a rule which tacitly assumes that lists of
hypotheticals are identified up to associativity:
$$\G_1,x:\tau,\G_2\vdash\Sigma$$
The first six rules of Figure~\ref{core} are the structural rules.
They are allow all other rules to be in a form where any
necessary assumptions appear as the leftmost child of the context.

\begin{figure}
\begin{minipage}{.48\linewidth}
\begin{align*}
\Sigma       ::= & \top\ |\ e:\tau^{\vec\eta}\ |\ \fo(\tau,\veta)
\\
\G           ::= & \Sigma\ |\ \ctree{\G}{\G}
\\
\eta         ::= & \text{\ level name}
\\
\vec \eta    ::= & \cdot\ |\ \eta, \vec\eta
\end{align*}
\end{minipage}
\begin{minipage}{.48\linewidth}
\begin{align*}
e           ::= & x\ |\ \lambda x.e\ |\ e[{\vec e}] |\ \code{e}\ |\ \ttt{\tild}e
\\
\vec e       ::= & \cdot\ |\ e, \vec e
\\
x            ::= & \text{\ expression variable}
\\
\tau         ::= & \tau\to\tau\ |\ \code{\tau^\eta}
\end{align*}
\end{minipage}

\caption{
Grammar for a simple multi-stage language.
\label{grammar}
}
\end{figure}

\begin{figure}
\setlength{\extrarowheight}{0.2mm}
\begin{tabular*}{30cm}{|p{11mm}|Mr@{}MlMcMl@{\hspace{3.9mm}}|}
\hline\vgap{2mm}
\text{RULE} & \text{SYNTAX} & & & \text{SEMANTICS}
\\\vgap{2mm}\hline\hline\vgap{\rospace}
\RN{Assoc}     & {\ctree{\G_1}{(\ctree{\G_2}{\G_3})}}\vdash & \Sigma  & = & \Delta
\\ \cline{2,3} & {\ctree{(\ctree{\G_1}{\G_2})}{\G_3}}\vdash & \Sigma  & = & \ttt{assoc}\ \>\ \Uuc{\Delta}
\interrow
\RN{Cossa}    & {\ctree{(\ctree{\G_1}{\G_2})}{\G_3}}\vdash & \Sigma  & = & \Delta
\\ \cline{2,3} & {\ctree{\G_1}{(\ctree{\G_2}{\G_3})}}\vdash & \Sigma  & = & \ttt{cossa}\ \>\ \Uuc{\Delta}
\interrow
\RN{Exch}      & {\ctree{\G_1}{\G_2}}\vdash & \Sigma  & = & \Delta
\\ \cline{2,3} & {\ctree{\G_2}{\G_1}}\vdash & \Sigma  & = & \ttt{swap}\ \>\ \Uuc{\Delta}
\interrow
\RN{Exch2}     & {\ctree{(\ctree{\G_1}{\G_2})}{\G_3}}\vdash & \Sigma  & = & \Delta
\\ \cline{2,3} & {\ctree{(\ctree{\G_2}{\G_1})}{\G_3}}\vdash & \Sigma  & = & (\ttt{first}\ \ttt{swap})\ \>\ \Uuc{\Delta}
\interrow
\RN{Weak}      & {\G_1}\vdash               & \Sigma  & = & \Delta 
\\ \cline{2,3} & {\ctree{\G_1}{\G_2}}\vdash & \Sigma  & = & \ttt{drop}\ \>\ \Uuc{\Delta} 
\interrow
\RN{Cont}      & {{\ctree{\G_1}{\G_1}}}\vdash & \Sigma  & = & \Delta
\\ \cline{2,3} & {{\G_1}}\vdash               & \Sigma  & = & \ttt{copy}\ \>\ \Uuc{\Delta}
\\\vgap{\rospace}\hline
\end{tabular*}
\vspace{-1mm}
\\
\begin{tabular*}{30cm}{|p{11mm}|Mr@{}Ml@{\hspace{2mm}}Mc@{\hspace{2mm}}Ml@{\hspace{2mm}}|}
\vgap{\rospace}
\RN{FC}        & \multicolumn{2}{c}{$\fo(\tau,({\eta,\veta}))$}    &   &
\\ \cline{2,3} & \multicolumn{2}{c}{$\fo(\code{\tau^\eta},\veta)$} &   &
\interrow
\RN{Var}       &  {}  & & &
\\ \cline{2,3} &  {x:\tau^\veta}\vdash  & x:\tau^\veta & = & \ttt{id}
\interrow
               &                                      & \fo(\tau_x,\veta)                  &   &
\\\RN{Lam}     &  {\ctree{x:\tau_x^\veta}{\G}}\vdash  & e:\tau^\veta                       & = & \Delta
\\ \cline{2,3} &  {\G}\vdash                          & \lambda x.e:(\tau_x{\to}\tau)^\veta  & = & \Uuc{\Delta}
\interrow
               &              & \fo(\tau,\veta)       &   &
\\\RN{App$_0$} &  {\G}\vdash  & e : \tau^\veta        & = & \Delta
\\ \cline{2,3} &  {\G}\vdash  & e [\cdot]: \tau^\veta & = & \Uuc{\Delta}
\interrow
                   &                                            & \fo(\tau_0,\veta)             &   &
\\                 &  \G_x                               \vdash & e_x:(\tau_0\to\tau_x)^\veta   & = & \Delta_x
\\                 &  \G_0                               \vdash & e_0:\tau_0^\veta              & = & \Delta_0
\\\RN{App$_{n{+}1}$} &  \ctree{x:\tau_x^\veta}{\G_e}       \vdash & x[\vec e]:\tau^\veta          & = & \Delta_1
\\ \cline{2,3}     &  \ctree{\G_x}{(\ctree{\G_0}{\G_e})} \vdash & e_x [e_0,\vec e] : \tau^\veta & = &
        \first\ \Uuc{\Delta_0}\\
& & & & \> \\
& & & & \second\ \Uuc{\Delta_1} \\
& & & & \> \\
& & & & \Uuc{\Delta_x}
\interrow
\RN{Brak}      &  {\G}\vdash  & e:\tau^{\eta,\veta}                & = & \Delta
\\ \cline{2,3} &  {\G}\vdash  & \code{e}:\code{\tau^{\eta}}^\veta  & = & \Uuc{\Delta}
\interrow
\RN{Esc}       &  {\G}\vdash  & e:\code{\tau^{\eta}}^\veta     & = & \Delta
\\ \cline{2,3} &  {\G}\vdash  & \ttt{\tild}e:\tau^{\eta,\veta} & = & \Uuc{\Delta}
\\\vgap{\rospace}\hline
\end{tabular*}

\vspace{3mm}

\nocaptionrule
\caption{
Typing rules for a simple multi-stage language, along with a
translation into generalized arrows.  The rules and translations are
rendered in the rule/syntax/semantics table style of \cite[Tables
  3,5,9]{Moggi:NotionsOfComputation}.  Note that contexts are
represented as a binary tree rather than a list.  An explanation of
the rules can be found in Section~\ref{explain-rules}.
\label{core}
}
\end{figure}

\begin{lemma}[Permutation of Contexts]
If there is a proof terminating in the judgement
$$\trfrac{\vdots}{\G_1\vdash\Sigma_1}$$ and some proposition $\Sigma_2$ appears as a leaf
of $\G_1$, then there is a proof terminating in the judgement
$$\trfrac{\vdots}{\ctree{\Sigma_2}{\G_2}\vdash\Sigma_1}$$ where the leaves
of $\ctree{\Sigma_2}{\G_2}$ are a permutation of the leaves of $\G_1$.
Furthermore, there is an algorithm for transforming the first proof tree
into the second.
\begin{proof}
in {\tt permutation\_of\_contexts} in {\tt GArrow.v}
\end{proof}
\end{lemma}

\subsection{Typing Rules for Staging Annotations}
\label{explain-rules}

The grammar for a simple multi-stage language can be found in
Figure~\ref{grammar}; the corresponding typing rules are in
Figure~\ref{core}.

\begin{remark}
Special attention should be paid to the superscripts used to denote
levels; a proposition $e:\tau^{\vec\eta}$ attributes a type $\tau$ to
an expression $e$ at a named level $\vec\eta$; the named level
$\vec\eta$ is part of the proposition, not the type.  Named levels do
not appear as part of types except the code type $\code{\tau^\eta}$,
which include exactly one level as part of the type; this level is
written {\it inside} the code-brackets.  The mnemonic justification
for this choice of syntax can be seen in the typing rules for
{\textsf{Brak}} and {\textsf{Esc}}.
\end{remark}

The first nonstructural rule, $\tsf{FC}$, distinguishes
types inhabited by {\it first class} values -- those that can be
arguments or return values of functions.  Because
$\fo(\tau{\to}\tau,\veta)$ is underivable without additional rules,
the type system as shown will prohibit first-class functions.
However, this restriction can easily be lifted by simply adding
another typing rule:
$$\trfrac{
\begin{trgather}
\fo(\tau_1,\veta) \\
\fo(\tau_2,\veta)
\end{trgather}
}{
\fo(\tau_1{\to}\tau_2,\veta)
}
$$
The next two rules are the variable ($\tsf{Var}$) and
abstraction ($\tsf{Lam}$) rules.  Note that the {\textsf{Var}} rule is
applicable only when the context contains {\it exactly} the assumption
needed and no others.  Any extraneous context elements must be
explicitly removed using {\textsf{Weak}}; this will be significant in
Section~\ref{affine-linear-ordered} which explores the possibility of
removing the {\textsf{Weak}} rule.  The {\textsf{Lam}} rule is
standard, save for the additional $\fo(\tau_x,\veta)$ hypothesis; this
ensures that abstractions over non-first-class values may not be formed.

The ${\tsf{App}}_0$ and ${\tsf{App}}_{n+1}$ provide for $n$-ary
function application via the $e[\vec e]$ production in the grammar.
After typechecking is complete, this $n$-ary application can be
syntactically expanded into $n$ instances of (curried) $1$-ary
application -- for example, $e[e_1, e_2, e_3, \cdot]$ becomes $(((e
e_1) e_2) e_3)$.  However, by having syntactic indication of the
application arity available {\it at typechecking time} the type system
can determine if a function application is {\it fully saturated}.
This is achieved via the $\fo(\tau,\veta)$ hypothesis in
${\tsf{App}}_0$, which prevents any function application from
producing a non-first-class value via unsaturated application.

The ${\tsf{App}}_{n+1}$ rule handles $n$-ary application for
$n{\geq}1$.  The first hypothesis is standard; the second ensures that
a function is never applied to a non-first-class value; the third is
standard and the fourth can be thought of as a recursive appeal to
${\tsf{App}}_n$.  Note that this rule does not assume that
the three subderivations take place under the same context.  In fact,
they must take place under separate contexts; this will matter if
{\textsf{Contr}} is removed.

The $\tsf{Brak}$ and $\tsf{Esc}$ rules are standard, copied from
\cite{Taha:ClassifierInference}.  Briefly, they prevent one piece of
code from being spliced into another using the {\tt{\tild}}$e$
construct unless both pieces of code are of the same depth (number of
surrounding brackets minus number of surrounding escapes is the same)
and their level names are the same.  The latter point will matter once
a type is introduced for {\it closed code} in
Section~\ref{eval-primitive}.

\section{The Translation}
\label{the-translation}

The translation from multi-stage programs to generalized arrows is
given by the rightmost column of Figure~\ref{core}, and is formalized
by the function {\tt translate} in {\tt GArrow.v}.  Note that the
translation operates on {\it proofs of well-typedness} rather than
expressions.

\begin{figure}
\begin{verbatim}
Definition pow : E V :=
 letrec pow := \\ n => \\ x =>
      If   (Eeq V) [ `n ; (Ezero V) ]
      Then <[Eone V]>
      Else <[(Emult V)[ ~~`x ;
              (~~ ((`pow) [ (Eminus V)[ `n ;
                 (Eone V)] ; `x ])) ] ]>
  in `pow.

Eval compute in (translate (pow_hastype _ n)).
letrec x := \\ x0 => \\ x1 =>
   If (first (`x0)
       >>> second ((first ga_true >>> second id)
                   >>> id))
       >>> ga_true
   Then ga_true
   Else (copy >>> (first copy >>> (swap >>>
        ga_true [`x1; copy >>> (first copy >>>
        (swap >>> (drop >>> id) [(first ((first
        (`x0) >>> second ((first ga_true >>>
        second id) >>> id)) >>> ga_true) >>>
        second ((first (`x1) >>> second id) >>>
        id)) >>> (`x); drop >>> id]))])))
   in (`x)
\end{verbatim}
\caption{ The {\tt pow} function's abstract syntax tree and the result
  of running the {\tt translate} procedure corresponding to the
  rightmost column of Figure~\ref{core} on it.  Note that the
  resulting abstract syntax tree does not contain any brackets or
  escapes; they have all been translated to equivalent
  \GArrow\ operations.
\label{translated-term}}
\end{figure}

The accompanying Coq formalization in {\tt GArrow.v} includes an
inductive type representing each of the productions in
Figure~\ref{grammar}, using a PHOAS \cite{AdamC:PHOAS} representation
for expressions.  Also included is an inductive type {\tt HasType} of
typing derivations under the rules of Figure~\ref{core}, and a
procedure {\tt translate}, which produces a \GArrow\ expression by
structural recursion on a {\tt HasType} proof.  An abstract syntax
tree for the {\tt pow} function is also included, and a corresponding
{\tt HasType} for it.  The result of applying the translation
procedure to a proof tht the {\tt pow} function is well-typed can be
found in Figure~\ref{translated-term}.

\begin{remark}
The fact that the translation operates on proofs rather than abstract
syntax trees has two curious practical consequences in the
accompanying {\tt GArrow.v}.  The first is that {\tt HasType} must
belong to {\tt Set} rather than {\tt Prop}, because although its inhabitants are
proofs their identities are not irrelevant.  The second is that the
unpleasant work of using the structural rules to re-arrange contexts
is easily automated using tacticals and the {\tt Ltac} scripting
language\footnote{This turned out to be far easier than expected}.
\end{remark}

The {\tt GArrow.v} formalization covers all material up to this point;
the remaining material is not included in the machine-checked portion
of this paper except where explicitly stated otherwise.

The remaining subsections will investigate possible object language
features which might be added, and the corresponding translation of
each feature into generalized arrows.  Each of the following
subsections is completely independent of the others; any combination
of the rule sets can be unioned with the rule set of Figure~\ref{core}
to produce an object language with that specific combination of
features.

\subsection{Recursive Let Bindings in Specific Stages}

Figure~\ref{types-letrec} gives syntax, typing rules, and translation
rules for the ability to permit recursion at specific levels and
types.  Note that the predicate $\tsf{recOk}$ is parameterized over
both the level $\veta$ and the type $\tau_x$ where the recursion
occurs.  This can be useful for:
\begin{itemize}
\item Allowing recursion only at certain stages.  For example, only in
      the metalanguage by adding the rule with no hypotheses and
      $\tsf{recOk}(\tau,\cdot)$ as the conclusion.
\item Allowing recursion only at certain types.  For example, allowing
      recursively-defined functions but not recursively-defined ground
      values at level $\veta$ by adding the rule with no hypotheses
      and $\tsf{recOk}(\tau\to\tau,\veta)$ as the conclusion.
\end{itemize}

\begin{figure}
\begin{align*}
e      & ::= \ttt{let}\ x{\ttt{=}}e\ttt{\ in\ }e\ |\ \ldots \\
\Sigma & ::= \tsf{recOk}(\tau,\veta)\ |\ \ldots
\end{align*}

\setlength{\extrarowheight}{0.6mm}
\begin{tabular}{|Mr|Mr@{}Ml@{\hspace{2mm}}Mc@{\hspace{2mm}}Ml|}
\hline\vgap{2mm}
\text{RULE} & \text{SYNTAX} & & & \text{SEMANTICS}
\\\vgap{2mm}\hline\hline\vgap{\rospace}
               &                                            & \tsf{recOk}(\tau_x,\veta)  & &
\\             &     \ctree{\ x{:}\tau_x^\veta\ }{\G_x}\vdash & e_x:\tau_x^\veta  & = & \Delta_x
\\\RN{Rec}     &     \ctree{\ x{:}\tau_x^\veta\ }{\G_e}\vdash & e:\tau^\veta      & = & \Delta_e
\\ \cline{2,3} &
   \ctree{\G_x}{\G_e}\vdash &
   \setlength{\extrarowheight}{-2mm}
   \begin{array}[t]{l}
   \ttt{let\ }x{\ttt{=}}e_x \\
   \ttt{in\ }e
   \end{array}{:}\tau^\veta
   & =
  &
  \setlength{\extrarowheight}{-2mm}
  \begin{array}[t]{l}
  \ttt{first}\ (\\
  \hspace{0.25cm}\ttt{loop}\ ( \\
  \hspace{0.50cm}\Uuc{\Delta_x} \\
  \hspace{0.50cm}\>\ \ttt{copy})) \\
  \>\ \Uuc{\Delta_e}
  \end{array}
\\\vgap{\rospace}\hline
\end{tabular}

\vspace{2mm}
\begin{verbatim}
  Class GArrowLoop ((**):Set->Set->Set)
                   ((~>):Set->Set->Set)
                   (  ga:GArrow (**) (~>)) :=
      loop : (a**c~>b**c) -> (a~>b)
\end{verbatim}
\caption{ Typing Rules for Recursive {\tt let} at Specific Stages.
  Assumes additional judgements for those stages at which recursive
  let-bindings are permitted.
\label{types-letrec}}
\end{figure}

\begin{figure}
\begin{align*}
\ttt{loop}\ (\first\ h\ \>\ f)
  &= h\ \>\ \ttt{loop}\ f
\\
\ttt{loop}\ (f\ \>\ \first\ h)
  &= \ttt{loop}\ f\ \>\ h
\\
\ttt{loop}\ (\ttt{loop}\ f)
  &= \ttt{loop}\ (\tt{cossa}\ \>\ f\ \>\ \ttt{assoc})
\\
\ttt{second}\ (\ttt{loop}\ f)
  &= \ttt{loop}\ (\tt{assoc}\ \>\ \color{red}\ttt{second}\color{black}\ f\ \>\ \ttt{cossa})
\end{align*}
\caption{Laws for the {\tt loop} function.\color{red}\ These follow the laws of 
\cite[Figure 7]{Paterson:NewNotationForArrows}, with ``Extension'' and ``Sliding'' omitted.
\label{loop-laws}}
\end{figure}

If recursion is to be used at any stage other than the first, it is
necessary for the {\tt GArrow} to also be a {\tt GArrowLoop} and
implement the {\tt loop} function of Figure~\ref{types-letrec}.  This
operation must satisfy the laws shown in Figure~\ref{loop-laws},
adapted from \cite[Figure 7]{Paterson:NewNotationForArrows}.  These
axioms first arose in work on traces on categories
\cite{StreetJoyalVerity:TracedMonoidal}, and were first applied to
functional programming in the context of value-recursive monads
\cite{Erkok:RecursiveMonadicBindings}.

\subsection{Booleans and Branching}
\label{bool-and-branch}

Figure~\ref{types-bool} gives grammar, typing rules, and translation
rules for boolean values and branching.  Note again that the
conditional and branches of the {\tt if} construct are typed under
disjoint pieces of the combined $\ctree{\G_i}{\G}$ context rather than
under a shared context.

\begin{figure}
\begin{align*}
\tau & ::= \ttt{bool}\ |\ \ldots \\
e    & ::= \ttt{true}\ |\ \ttt{false}\ |\ \ttt{if\ }e\ttt{\ then\ }e\ttt{\ else\ }e\ |\ \ldots 
\end{align*}

\setlength{\extrarowheight}{0.6mm}
\begin{tabular}[C]{|Mr|Mr@{}Ml@{\hspace{2mm}}Mc@{\hspace{2mm}}Ml|}
\hline\vgap{2mm}
\text{RULE} & \text{SYNTAX} & & & \text{SEMANTICS}
\\\vgap{2mm}\hline\hline\vgap{\rospace}
\RN{Bool}      &  & & &
\\ \cline{2,3} &  \multicolumn{2}{c}{$\fo(\ttt{bool},\veta)$} & & 
\interrow
\RN{True}      &  & & &
\\ \cline{2,3} &  \top\vdash & \ttt{\ true}:\ttt{bool}^\veta & & 
\interrow
\RN{False}     &  & & &
\\ \cline{2,3} &  \top\vdash & \ttt{\ false}:\ttt{bool}^\veta & & 
\interrow
               &  \G_i\vdash   & e_i:\ttt{bool}^\veta & = & \Delta_i
\\             &  \G  \vdash & e_t:\tau^\veta     & = & \Delta_t
\\\RN{If}      &  \G  \vdash & e_e:\tau^\veta     & = & \Delta_e
\\ \cline{2,3} &  \ctree{\G_i}{\G}\vdash
               &  \setlength{\extrarowheight}{-2mm}
                  \begin{array}[t]{l}
                  \ttt{if\ }e_i \\
                  \ttt{then\ }e_t \\
                  \ttt{else\ }e_e
                  \end{array}{:}\tau^\veta & = &
 \begin{array}[t]{l}
 (\ttt{first\ }\Uuc{\Delta_i})\ \>\\
 (\ttt{branch\ }\Uuc{\Delta_t}\ \Uuc{\Delta_e})
 \end{array}
\\\vgap{\rospace}\hline
\end{tabular}

\vspace{2mm}
\begin{verbatim}
  Class GArrowBool ((**):Set->Set->Set)
                   ((~>):Set->Set->Set)
                   (  ga:GArrow (**) (~>)) :=
    branch : (a~>b) -> (a~>b) -> ((bool**a)~>b)
\end{verbatim}
\caption{
Typing Rules for booleans.
\label{types-bool}}
\end{figure}

\subsection{Cross-Stage Persistence}

Figure~\ref{types-reification} gives the rules for cross-stage
persistence (CSP).  CSP is permitted only for fully-normalized values
belonging to a non-function (ground) type; these types are
distinguished by the {\textsf{reifiable}$(\tau,\veta)$} judgement.
Appropriate inference rules must be added for whatever kinds of types
(primitives, products, coproducts, etc) are in the system to ensure
that {\textsf{reifiable}}$(\tau,\veta)$ is derivable for those types at
which it is appropriate.

\begin{figure}
\begin{align*}
e      & ::= \ttt{\%}e\ |\ \ldots \\
\Sigma & ::= \tsf{reifiable}(\tau,(\eta,\veta))\ |\ \ldots
\end{align*}
\setlength{\extrarowheight}{0.6mm}
\begin{tabular}[C]{|Mr|Mr@{}Ml@{\hspace{2mm}}Mc@{\hspace{2mm}}Ml|}
\hline\vgap{2mm}
\text{RULE} & \text{SYNTAX} & & & \text{SEMANTICS}
\\\vgap{2mm}\hline\hline\vgap{\rospace}
               &              & \tsf{reifiable}(\tau,\veta)  &   &
\\\RN{CSP}     &     \G\vdash & e:\tau^\veta                 &   &
\\ \cline{2,3} &     \G\vdash & \ttt{\%}e:\tau^{\eta,\veta}  & = & \ttt{reify\ }e
\\\vgap{\rospace}\hline
\end{tabular}

\vspace{2mm}
\begin{verbatim}
Class GArrowReify ((**):Set->Set->Set)
                  ((~>):Set->Set->Set)
                  (  ga:GArrow (**) (~>)) :=
  reify : (a->b) -> (a~>b)
\end{verbatim}

\color{red}
\begin{verbatim}
  reify_extensional :
   forall {a}{b}{f:a->b}{g},
    (forall x, (f x)=(g x))
    -> (reify f)~~(reify g)
\end{verbatim}
\color{black}

\caption{Typing rules for cross-stage persistence (CSP).\label{types-reification}}
\end{figure}

\subsection{Product Types in the Object Language}

Figure~\ref{product-types} gives rules for product types.

\color{red}
The laws given are exactly those needed to ensure that the {\tt <*>}
operator induces a {\it finite product}
(Definition~\ref{finite-product-category}) structure with
$!X=\ttt{drop}$ and $\Delta_X=\ttt{delta}$.  FIXME: should the GArrow
itself choose {\tt unit}?
\color{black}

\begin{remark}
Note that {\tt **} and $\otimes$ are not the same.  The {\tt **}
operator represents {\it contexts}, which are not first-class in the
object language.  The $\otimes$ operator represents products,
which {\it are} first-class in the object language.
\end{remark}

\Arrows\ do not make the distinction above, which is a source of
limitations.  For example, an \Arrow\ for stream processors does not
distinguish between a {\it pair of streams} and a {\it stream of
  pairs}; both are {\tt a*b{\tild}>c*d} (which is a retract of
{\tt (a{\tild}>c)*(b{\tild}>d)} in the absence of side effects).  With \GArrows\ {\it pairs
  of streams} have type {\tt a**b{\tild}>c**d} and {\it streams of
  pairs} have type {\tt a${\otimes}$b{\tild}>c${\otimes}$d}.  In a
{\it synchronous dataflow} environment these two concepts coincide;
this explains why all existing literature on using \Arrows\ for stream
processing \cite{PembeciNilssonHager:ArrowizedFRP,Hughes:GeneralisingMonadsToArrows}
and digital circuits
\cite{Paterson:NewNotationForArrows,Hughes:ProgrammingWithArrows} applies only to synchronous
environments.  Attempts to create \Arrows\ for unrestricted Petri Nets
\cite{Petri:Thesis} are impeded by this limitation.  The need to have
distinct types for ``stream of pairs'' and ``pair of streams'' led the
Fudgets library to co-opt the {\it coproduct} structure of the underlying
type system to represent pairs of streams, which explains the anomoly
that Paterson notes \cite[Section 5.1]{Paterson:NewNotationForArrows}
in the type of the Fudgets {\tt loop} function
\cite{CarlssonHallgren:Fudgets}.

\subsection{Coproduct Types in the Object Language}

Figure~\ref{coproduct-types} gives the rules for coproduct types.  The
{\tt branch} and {\tt bool} of Section~\ref{bool-and-branch} can be
seen as a restricted form of {\tt c\_merge} and {\tt <+>}.

\begin{figure*}
\begin{minipage}[b]{.48\linewidth}
\begin{align*}
\tau ::=& \tau\otimes\tau\ |\ \ldots \\
e    ::=& \ttt{fst}\ e\ |\ \ttt{snd}\ e\ |\ \langle{e,e}\rangle\ |\ \ldots
\end{align*}

\setlength{\extrarowheight}{0.6mm}
\begin{tabular}[C]{|Mr|Mr@{}Ml@{\hspace{2mm}}Mc@{\hspace{2mm}}Ml|}
\hline\vgap{2mm}
\text{RULE} & \text{SYNTAX} & & & \text{SEMANTICS}
\\\vgap{2mm}\hline\hline\vgap{\rospace}
                            & \multicolumn{2}{c}{$\fo(\tau_1,\veta)$}  &   &
\\\RN{FC$_\text{prod}$}     & \multicolumn{2}{c}{$\fo(\tau_2,\veta)$}  &   &
\\ \cline{2,3}              & \multicolumn{2}{c}{$\fo(\tau_1{\otimes}\tau_2,\veta)$}  &   &
\interrow
  \RN{Fst}     &     \G\vdash & e:(\tau_1\otimes\tau_2)^\veta &   & \Delta
\\ \cline{2,3} &     \G\vdash & \ttt{fst}\ e:\tau_1^\veta     & = & 
\hspace{-3mm}
\color{red}
  \setlength{\extrarowheight}{-2mm}
  \begin{array}[t]{l}
  \ttt{lift(id**drop)}\\
  \>\ \ttt{iso1}\ \>\ \Uuc{\Delta}
  \end{array}
\color{black}
\interrow
  \RN{Snd}     &     \G\vdash & e:(\tau_1\otimes\tau_2)^\veta &   & \Delta
\\ \cline{2,3} &     \G\vdash & \ttt{snd}\ e:\tau_2^\veta & = &
\hspace{-3mm}
\color{red}
  \setlength{\extrarowheight}{-2mm}
  \begin{array}[t]{l}
  \ttt{lift(drop**id)}\\
  \>\ \ttt{iso2}\ \>\ \Uuc{\Delta}
  \end{array}
\color{black}
\interrow
               &                \G_1\vdash & e_1:\tau_1^\veta &   & \Delta_1
\\\RN{Prod}    &                \G_2\vdash & e_2:\tau_2^\veta &   & \Delta_2
\\ \cline{2,3} & \ctree{\G_1}{\G_2} \vdash & {\langle}e_1{,}e_2{\rangle}{:}(\tau_1{\otimes}\tau_2)^\veta & = &
  \setlength{\extrarowheight}{-2mm}
\color{red}
  \begin{array}[t]{l}
  \ttt{lift}\ (\\
  \ \ \first\ \Uuc{\Delta_1}\\
  \ \ \> \\
  \ \ \second\ \Uuc{\Delta_2})
  \end{array}
\color{black}
\\\vgap{\rospace}\hline
\end{tabular}

\begin{verbatim}
Class GArrowProd (g:GArrow G)
                 ((<*>):Set->Set->Set) :=
\end{verbatim}
\color{red}
\begin{verbatim}
  unit  : Set

  delta : a ~> a<*>a
  iso1  : a<*>unit ~> a
  iso2  : unit<*>a ~> a
  lift  : (a**b)~>(c**d) -> (a<*>b)~>(c<*>d)
  id ~~ delta >>> (lift (id   *** drop)) >>> iso1
  id ~~ delta >>> (lift (drop *** id  )) >>> iso2
\end{verbatim}
\color{black}

\vfill
\caption{Product Types \label{product-types}}
\end{minipage}
\begin{minipage}[b]{.48\linewidth}

\begin{align*}
\tau ::=& \tau\oplus\tau\ |\ \ldots \\
e    ::=& \ttt{inl}\ e\ |\ \ttt{inr}\ e\ |\ \ttt{case\ }e\ttt{\ of\ |\ L\ }x\ttt{\ ->\ }e\ttt{\ |\ R\ }x\ttt{\ ->\ }e\ |\ \ldots
\end{align*}

\setlength{\extrarowheight}{0.6mm}
\begin{tabular}[C]{|Mr|Mr@{}Ml@{\hspace{2mm}}Mc@{\hspace{2mm}}Ml@{}|}
\hline\vgap{2mm}
\text{RULE} & \text{SYNTAX} & & & \text{SEMANTICS}
\\\vgap{2mm}\hline\hline\vgap{\rospacex}
                              & \multicolumn{2}{c}{$\fo(\tau_1,\veta)$}  &   &
\\\RN{FC$_\text{coprod}$}     & \multicolumn{2}{c}{$\fo(\tau_2,\veta)$}  &   &
\\ \cline{2,3}                & \multicolumn{2}{c}{$\fo(\tau_1{\oplus}\tau_2,\veta)$}  &   &
\interrowx
  \RN{InL}     &     \G\vdash & e:\tau_1^\veta & = & \Delta
\\ \cline{2,3} &     \G\vdash & \ttt{inl\ }e:(\tau_1\oplus\tau_2)^\veta  & = &
\hspace{-3mm}
\color{red}
  \setlength{\extrarowheight}{-2mm}
  \begin{array}[t]{l}
  \ttt{iso1}\ \>\\
  \ttt{lift(id**codrop)}\\
  \>\ \Uuc{\Delta}
  \end{array}
\color{black}
\interrowx
  \RN{InR}     &     \G\vdash & e:\tau_2^\veta & = & \Delta
\\ \cline{2,3} &     \G\vdash & \ttt{inr\ }e:(\tau_1\oplus\tau_2)^\veta  & = &
\hspace{-3mm}
\color{red}
  \setlength{\extrarowheight}{-2mm}
  \begin{array}[t]{l}
  \ttt{iso2}\ \>\\
  \ttt{lift(codrop**id)}\\
  \>\ \Uuc{\Delta}
  \end{array}
\color{black}
\interrowx
               &                         \G_0\vdash & e_0:(\tau_1{\oplus}\tau_2)^\veta & = & \Delta_0
\\             &   \ctree{\G}{x{:}\tau_1^\veta}\vdash & e_1:\tau^\veta                 & = & \Delta_1
\\\RN{CP}  &   \ctree{\G}{x{:}\tau_2^\veta}\vdash & e_2:\tau^\veta                 & = & \Delta_2
\\ \cline{2,3} & \ctree{\G_0}{\G}{\vdash} &
  \setlength{\extrarowheight}{-7mm}
  \begin{array}[t]{l}
   \ttt{case\ }e_0\ttt{\ of\ }\\
   \hspace{0.15cm}\ttt{|\ L\ }x\ttt{->\ }e_1 \\
   \hspace{0.15cm}\ttt{|\ R\ }x\ttt{->\ }e_2
   \end{array}{:}\tau^\veta
  & = &
  \setlength{\extrarowheight}{-2mm}
\color{red}
  \begin{array}[t]{l}
  \ttt{lift}\ (\\
  \ \ \first\ \Uuc{\Delta_1}\ \>\\
  \ \ \second\ \Uuc{\Delta_2})\ \>\ \ttt{codelta}
  \end{array}
\color{black}
\\\vgap{\rospacex}\hline
\end{tabular}

\begin{verbatim}
Class GArrowCoprod (g:GArrow G)
                   ((<+>):Set->Set->Set) :=
\end{verbatim}
\color{red}
\begin{verbatim}
  void    : Set            (* the uninhabited type *)
  codrop  : void ~> a
  codelta : a<+>a ~> a
  iso1    : a ~> a<+>void
  iso2    : a ~> void<+>a
  lift    : (a**b)~>(c**d) -> (a<+>b)~>(c<+>d)
  id ~~ iso1 >>> (lift (id   *** codrop)) >>> codelta
  id ~~ iso2 >>> (lift (codrop *** id  )) >>> codelta
\end{verbatim}
\color{black}

\caption{Coproduct Types\label{coproduct-types}}
\end{minipage}
\end{figure*}

\subsection{Affine, Linear, and Ordered Types in the Object Language}
\label{affine-linear-ordered}

Affine types in the object language can be modeled by omitting {\tt
  copy} (eliminating the $\tsf{Cont}$ rule); linear types can be
simulated by omitting {\tt copy} and {\tt drop} (eliminating the
$\tsf{Weak}$ rule).  Ordered linear types
\cite{PolakowPfenning:OrderedLogic} can be imitated by omitting {\tt
  swap} (eliminating the $\tsf{Exch}$ rule).

\begin{remark}
If {\tt swap} is omitted, the definition of {\tt cossa} is no longer
redundant, and it must be defined separately.
\end{remark}

Typechecking and type inference for affine, linear, and ordered types
is a complex topic.  This paper does not attempt to address these
questions; it takes the finished typing derivation as a starting point
for the translation procedure.

\subsection{The {\tt eval} Primitive}
\label{eval-primitive}

The rules for {\tt eval} (also called {\tt run}) can be found in
Figure~\ref{rules-eval}.  The {\tt eval} primitive can only be used
safely on {\it closed code}; the {\tt open} and {\tt close} primitives
are needed to mark such regions \cite{Taha:ClassifierInference}.

The {\tt GArrowEval} class, which has a {\tt Prop} index but no
methods, has a close relationship to Haskell's {\tt runST}, the {\it
  strict state monad} \cite{LaunchburyJones:RunST} which has rank-2
type:

\begin{verbatim}
  runST :: (forall s. ST s a) -> a
\end{verbatim}

The {\tt runST} function has this type in order to ensure that values
returned by {\tt runST} do not contain ``dangling references'' to the
state index {\tt s}.  This effect is achieved by taking advantage of
the fact that the introduction rule for $e:(\forall\alpha)\tau$
requires that $\alpha$ not appear in the type environment -- it is a
closedness condition, albeit upon types rather than values (no matter:
parametricity supplies the linkage).  This closedness condition on
types and values closely paralells the closedness conditions in the
hypothesis of the $\tsf{Close}$ rule, which must be applied before
{\tt eval}.

\begin{theorem}
The translation converts staged values of {\it closed} type
$\code{\tau^\Box}$ to expressions of a rank-2 type
parametric over the \GArrow\ instance.
\begin{proof}
in {\tt translation\_of\_closed\_code\_is\_parametric}\\ in {\tt GArrow.v}
\end{proof}
\end{theorem}

\begin{figure}
\begin{align*}
\tau & ::= \code{\tau^\Box}\ |\ \ldots \\
e    & ::= \ttt{open}\ e\ |\ \ttt{close}\ e\ |\ \ttt{eval}\ e\ |\ \ldots
\end{align*}

\setlength{\extrarowheight}{0.6mm}
\begin{tabular}[C]{|Mr|Mr@{}Ml@{\hspace{2mm}}Mc@{\hspace{2mm}}Ml|}
\hline\vgap{2mm}
\text{RULE} & \text{SYNTAX} & & & \text{SEMANTICS}
\\\vgap{2mm}\hline\hline\vgap{\rospace}
  \RN{Open}    &     \G\vdash & e:\code{\tau^\Box}^\veta                          & = & \Delta
\\ \cline{2,3} &     \G\vdash & \ttt{open}\ e:\code{\tau^{\eta'}}^\veta & = & \Uuc{\Delta}
\interrow
               & \multicolumn{2}{c}{$\eta'\notin\tsf{FV}(\G,\veta,\tau)$}         &   &
\\\RN{Close}   &     \G\vdash & e:{\code{\tau^{\eta'}}}^\veta                     & = & \Delta
\\ \cline{2,3} &     \G\vdash & \ttt{close}\ e:{\code{\tau^\Box}}^\veta & = & \Uuc{\Delta}
\interrow
  \RN{Eval}    &     \G\vdash & e:\code{\tau^\Box}^\veta                          & = & \Delta
\\ \cline{2,3} &     \G\vdash & \text{\ttt{eval}}\ e:\tau^\veta                   & = & \ttt{eval}\ \Uuc{\Delta}
\\\vgap{\rospace}\hline
\end{tabular}

\vspace{2mm}
\begin{verbatim}
Class GArrowEval ((**):Set->Set->Set)
                 ((~>):Set->Set->Set)
                 (  ga:GArrow (**) (~>)) :=
                 (idx:Prop) := { }.
eval : forall ((**):Set->Set->Set)
              ((~>):Set->Set->Set)
              (  ga:GArrow (**) (~>)),
    (forall (idx:Prop),
          (GArrowEval (**) (~>) ga idx) -> (a~>b))
    -> (a->b).
\end{verbatim}
\caption{Rules for {\tt eval}.\label{rules-eval}}
\end{figure}

\section{Examples}
\label{examples-section}

\subsection{Exponentiation of Natural Numbers}

It is now time to return to the example program, {\tt pow}, expressed
using staging annotations:

\begin{verbatim}
  pow n x =
    if n==0
    then <[ 1 ]>
    else <[ ~x * ~(pow (n-1) x) ]>
\end{verbatim}

\begin{theorem}
For any $\veta$, there exists a typing derivation using the rules of
Figures~\ref{core} and \ref{types-bool} for
$\Gamma\vdash\ttt{pow}:\ttt{Int->}\code{\ttt{Int}}\ttt{->}\code{\ttt{Int}}^\veta$
where $\Gamma$ contains suitable type assumptions for {\tt 0}, {\tt
  1}, {\tt (*)}, {\tt (-)}, and {\tt (==)}.
\begin{proof}
in {\tt pow\_hastype} in {\tt GArrow.v}
\end{proof}
\end{theorem}

\subsection{BiArrows}
\label{biarrows}

{\tt BiArrows} are meant to model \Arrows\ with a notion of {\it
  inversion}.  They were introduced in \cite{ASWEP:BiArrows} and
further examined in
\cite{JacobsHeunenHasuo:CategoricalSemanticsArrows}.  Briefly,

\begin{verbatim}
Class BiArrow ((~>):Set->Set->Set)
              (arrow:Arrow (~>)) :=

  biarr : (a->b) -> (b->a) -> (a~>b)
  inv   :   a~>b -> b~>a

  pf0   :   inv (biarr f f') ~~ biarr f' f
  pf1   :        inv (inv f) ~~ f
  pf2   :      inv (g >>> f) ~~ (inv f) >>> (inv g)
  pf3   :        inv (arr f) ~~ (arr swap)
  pf4   :      inv (first f) ~~ first (inv f)
\end{verbatim}

The {\tt BiArrow} class adds a new constructor {\tt biarr}, which is
to be used in place of {\tt arr}.  It takes a pair of functions which
are required to be mutual inverses.  The {\tt inv} function attempts
to invert a {\tt BiArrow}.

Types belonging the class {\tt BiArrow} consist of operations which
{\it might be} invertible.  Some {\tt BiArrow} values are actually not
invertible, so the {\tt inv} operation is only partial and may fail at
runtime.  The type system is not capable of ensuring that ``well-typed
programs cannot go wrong'' in this way.  Unfortunately there is no way
to fix this within the framework of \Arrows, because the \Arrow\ type
class requires that {\tt arr} be defined for arbitrary functions --
even those like {\tt fst} (the first projection of a tuple) which
cannot possibly have an inverse.  Moreover, the {\tt arr} function is
tightly woven in to the laws which prescribe the behavior of \Arrows,
so solving the problem is not as simple as replacing {\tt arr} with
{\tt biarr}.

However, one {\it can} create a \GArrow\ which preserves invertibility.
There are two possibilities, in fact:

\begin{itemize}
\item Realize the \GArrow\ {\tt drop} method using the {\it logging
      translation} of \cite[Section
      6]{MuHuTakeichi:InjectiveReversible}, which implements tuple
      projection by concealing the non-projected coordinates
      rather than discarding them entirely.

\item Declare a superclass of \GArrow\ which omits the {\tt drop}
      function.  This is not nearly as violent a change as attempting
      to remove {\tt arr} from \Arrow; the translation of
      Figure~\ref{core} remains intact for any derivation which does not
      use the $\tsf{Weak}$ rule.  As a result, object programs
      typeable under certain variants of linear logic remain
      translatable.
\end{itemize}

\subsection{Circuit Description}

Many researchers have investigated the use of functional programming
languages to describe hardware circuits
\cite{BCSS:Lava,Grundy:Reflect,Matthews:MicroprocessorInHawk,Sharp:SAFL,Sharp:Ruby}.
The allure is strong: combinational circuits and pure functions have
much in common.  However, in order to create usable circuits one must
allow for sharing and feedback, and this is where the similarities
end.

Pure functional languages which represent circuit nodes as first-class
language values must add an impurity, {\it observable sharing}
\cite{ClaessenSands:ObservableSharing}, to the language in order to
preserve sharing information and permit introspection on circuits with
feedback.  This impurity is incompatible with optimizations present in
many compilers for pure functional languages and considerably
complicates the semantics of the language.  The alternative is to
represent circuits using a value-recursive monad
\cite{Erkok:RecursiveMonadicBindings} or \Arrow; this avoids the
pitfalls of observable sharing but requires that circuits be
constructed in an object language which is completely different from
the functional metalanguage -- a choice which dilutes the benefits
sought.

With the translation from staging annotations to \GArrows, programmers
can write circuits {\it and circuit generators} with a single set of
binding, scoping, abstraction, and application mechanisms.

\section{Categorical Perspective}
\label{categorical-perspective}

The time has come to make good on the promise of the paper's subtitle.
Technically what will be exhibited in this section is an {\it
  equivalence} of categories, but -- like every equivalence -- this
will give an isomorphism of skeletons.

In addition to abstract theorems involving categories, most
subsections of this section will include an example involving a
category $\O$ whose objects are the types of some object programming
language (pick your favorite side-effect free language) and whose
morphisms are the functions of that language.

\begin{definition}[{\cite[Definition 2.7]{Awodey:CategoryTheory}}]
An object $1$ of a category $\C$ is the {\it terminal object} if there
is exactly one morphism into $1$ from every other object.  This
morphism will be written $!A:A{\to}1$.
\end{definition}

\begin{definition}[{\cite[3.2]{PowerRobinson:PremonoidalCategories}}]
A {\it binoidal category} is a category $\C$ given with a pair of
bifunctors $-{\ltimes}-:\C{\times}\C\to\C$ and
$-{\rtimes}-:\C{\times}\C\to\C$ such that for all objects $A,B$ of
$\C$ it is the case that $A{\ltimes}B=A{\rtimes}B$, which is also
written $A{\otimes}B$.
\end{definition}

\begin{definition}[{\cite[3.3]{PowerRobinson:PremonoidalCategories}}]
\label{def-central}
A morphism $f$ for which it is the case that
$f{\ltimes}g=f{\rtimes}g$ for all $g$ is called a {\it central}
morphism.
\end{definition}

Binoidal categories are generally used to model computations in which
{\it evaluation order} is significant.  The fact that the two
bifunctors agree on objects reflects the fact that type systems do not
track which coordinate of a tuple was computed first.  The fact that
the bifunctors may disagree on morphisms reflects the fact that
evaluating the left coordinate first may yield a different result than
evaluating the right coordinate first.  Central maps model
computations which are {\it pure} and therefore commute (in time) with all
others.  Note that for morphisms $f$ and $g$ the expression
$f{\otimes}g$ is not well-defined unless at least one of $f$ or $g$ is
central.

\begin{definition}[{\cite[3.5]{PowerRobinson:PremonoidalCategories}}]
A {\it premonoidal category} is a binoidal category with an object $I$
such that $A{\otimes}(B{\otimes}C) \cong (A{\otimes}B){\otimes}C$ and
$X{\otimes}I \cong X \cong I{\otimes}X$ for all objects $X$ subject to
the coherence conditions of \cite[p162]{MacLane:CWM}.  A {\it strict
  premonoidal category} is a premonoidal category in which the above
isomorphisms are identity maps.  A {\it premonoidal functor} is a
functor between premonoidal categories which preserves this structure.
\end{definition}

\begin{definition}
A {\it symmetric premonoidal category} is a category in which
$A{\otimes}B \cong B{\otimes}A$ and the mediating isomorphism is its
own inverse.
\end{definition}

\begin{definition}
A {\it monoidal category} is a premonoidal category in which every map
is central.
\end{definition}

Note that a category may be monoidal in more than one way: there may
be multiple bifunctors that satisfy the properties above.  For example
{\bf Sets}, the category of sets and functions, is monoidal under not
only cartesian product but disjoint union as well.  The same applies
to binoidality and premonoidality.

\begin{definition}
\label{finite-product-category}
A {\it finite product category} is a monoidal category \color{red}\ in
which $I=1$ is a terminal object along with a morphism
$\Delta_X:X{\to}X{\otimes}X$ for each object $X$ such that the
following diagram commutes:
\begin{center}
\begin{tikzpicture}[descr/.style={fill=white,inner sep=2.5pt}]
   \matrix (m) [matrix of math nodes, row sep=3em, column sep=3em] {
     1{\otimes}X & X{\otimes}X \\
     X           & 1{\otimes}X \\ };
   \path[->,font=\scriptsize] 
    (m-2-1) edge node[above,sloped] {$\Delta_X$} (m-1-2)
    (m-1-2) edge node[auto]         {$\tsf{id}_X \otimes !X$}  (m-2-2)
    (m-1-2) edge node[auto,swap]    {$!X \otimes \tsf{id}_X$}  (m-1-1)
    (m-1-1) edge node[auto,swap]    {$\cong$} (m-2-1)
    (m-2-2) edge node[auto]         {$\cong$} (m-2-1)
   ;
\end{tikzpicture}
\end{center}
FIXME: and is equal to the identity -- need another branch
\color{black}
A {\it finite product functor} is a functor between
finite product categories which preserves this structure.
\end{definition}

In a finite product category the monoidal functor will be written
$\times$ rather than $\otimes$ to emphasize this additional structure.
Note that $1$ is the $0$-ary product; zero is considered finite in
this paper.

\begin{definition}[{\cite[Definition B1.2.1(a)]{Johnstone:SketchesOfAnElephant}}]
\label{indexed-category}
For $\C$ a category, a {\it $\C$-indexed category} $\D^{(-)}$ assigns
a category $\D^A$ to each object $A$ of $\C$ and a functor
$\D^f:\D^X\to\D^Y$ to each morphism $f:X\to Y$ of $\C$ in such a way
that $\D^f\circ\D^g \cong \D^{g\circ f}$.  If $\C$ has a terminal
object $1$, then $\C\cong \D^1$.
\end{definition}

\begin{definition}[{\cite[Definition B1.2.1(b)]{Johnstone:SketchesOfAnElephant}}]
\label{indexed-functor}
An {\it $\C$-indexed functor} $F^{(-)}:\D^{(-)}\to\E^{(-)}$ assigns to each object
$A$ of $\C$ a functor $F^A:\D^A\to\E^A$ and to each morphism $f:X\to
Y$ a natural isomorphism $F^f:(F^Y\circ\D^f)\cong(\E^f\circ F^X)$
allowing the following diagram to commute up to isomorphism of functors:
\begin{center}
\begin{tikzpicture}[descr/.style={fill=white,inner sep=2.5pt}]
   \matrix (m) [matrix of math nodes, row sep=3em, column sep=3em] {
     \D^Y & \E^Y \\
     \D^X & \E^X \\ };
   \path[->,font=\scriptsize] 
    (m-2-2) edge node[auto,swap] {$\E^f$}  (m-1-2)
    (m-1-1) edge node[auto] {$F^Y$} (m-1-2)
    (m-2-1) edge node[auto,swap]      {$F^X$} (m-2-2)
    (m-2-1) edge node[auto]      {$\D^f$}  (m-1-1)
   ;
\end{tikzpicture}
\end{center}
\end{definition}

\begin{definition}
\label{exponential}
For a category $\C$ with monoidal bifunctor $(-){\otimes}(-)$, a
$\otimes$-{\it{exponential}} is a bifunctor $(-){\Rightarrow}(-)$ such
that for each object $B$ of $\C$, the functor $B{\Rightarrow}(-)$ is
right adjoint to the functor $(-){\otimes}B$.
\end{definition}

An $\otimes$-exponential induces the following isomorphism of Hom-sets:

\begin{center}
\begin{tabular}[c]{r@{}l}
$A{\otimes}$&$B\to C$ \\
\hline
\hline
$A \to$&$B \Rightarrow C$
\end{tabular}
\end{center}

\begin{definition}
A {\it cartesian closed category} is a finite product category with a
$\times$-exponential.
\end{definition}

\begin{remark}
The definition of {\it exponential} is usually stated in a form
specific to cartesian products.  The more general definition above
will allow investigation of exponentials over monoidal structure which
is not necessarily a cartesian product.
\end{remark}

\subsection{Polynomial Categories}

Most algebraists are familiar with the construction whereby one passes
from a ring $R$ to the ring $R[x]$ of polynomials with one
indeterminate and coefficients from $R$.  A similar construction is
possible with categories.

\begin{definition}[Provisional]
Given a category $\C$ with a terminal object $1$, and some object $B$ of
$\C$, let the {\it polynomial category over $\C$ in $B$}, written
$\C[x{:}B]$, be the free category obtained by adjoining to $\C$ a new
morphism $x:1{\to}B$ and closing under composition and products of
morphisms.  The morphisms of $\C[x{:}B]$ are called {\it polynomials over
  $\C$ in $B$}.  \cite[Definition 2.5]{Lambek:FunctionalCompleteness}
\end{definition}

Like the free group on a set, this ``free category obtained by
adjoining a new morphism'' can be understood intuitively as the
category including $x{:}1{\to}B$ while introducing as few new
morphisms and satisfying as few new identities as possible.  Terms
with free variables in them are best understood as morphisms in a
polynomial category, and variable-binding operators as functors from
the polynomial category back into the host category.  This gives some
semantic weight to the notion of a ``term definable in terms of some
hypothetical of type $B$'' -- these are exactly the morphisms of
$\C[x{:}B]$.

This paper will generally represent polynomial morphisms
(except for the indeterminate $x$) using lower-case letters with a
superscript, such as $f^B$, as a reminder that $f^B$ belongs to
$\C[x{:}B]$ rather than $\C$.

\begin{definition}[Provisional]
\label{weakening-functor}
The {\it weakening functor} of a category $\C$ assigns to each object
$B$ of $\C$ a functor $\C^{!B}:\C{\to}\C[x{:}B]$ from $\C$ to the
polynomial over $\C$ in $B$ such that $\C^{!B}$ is the inclusion
functor when $\C$ is regarded as a subcategory of $\C[x{:}B]$.
\end{definition}

\begin{remark}
\label{cartesian-remark}
If it happens that $\C$ is a finite product category, one can construct
$\C[x{:}B]$ and the weakening functor explicitly: the weakening
functor sends each object $A$ to $B{\times}A$ and each morphism $f$ to
$\tsf{Id}_B{\times}f$.  $\C[x{:}B]$ is the subcategory of $\C$ which
is the range of this functor.  However, if $\C$ has a weaker monoidal
structure (perhaps only premonoidal), or none at all, the notion of
polynomial category is not definable in this manner.
\end{remark}

A slightly more rigorous formulation, adapted from \cite[Remark
  2.6]{Lambek:FunctionalCompleteness}, can be given in terms of
indexed categories and universal properties:

\begin{definition}[Official]
\label{def-poly}
For $\C$ a category with a terminal object $1$, a {\it polynomial
  category} $\C[x{:}-]$ is a $\C$-indexed category such that for every
object $B$, functor $G{:}\C{\to}\D$ and $d{:}1{\to}G(B)$
there exists a unique functor $\subst{x}{d}(-){:}\C[x{:}B]{\to}\D$ such
that $\subst{x}{d}(x)=d$ and $\subst{x}{d}{\circ}\C^{!B}=G$.
\begin{center}
\begin{tikzpicture}[descr/.style={fill=white,inner sep=2.5pt}]
   \matrix (m) [matrix of math nodes, row sep=3em, column sep=3em] {
     \C[x{:}B] &      \\
     \C        & \D   \\ };
   \path[->,font=\scriptsize] 
    (m-2-1) edge node[auto] {$\C^{!B}$}               (m-1-1)
    (m-1-1) edge node[auto] {$\exists \subst{x}{d}(-)$}  (m-2-2)
    (m-2-1) edge node[auto,swap]      {$\forall G$}             (m-2-2)
   ;
\end{tikzpicture}
\end{center}
The
functor $\C^{!B}$ is called the {\it weakening functor} at $B$.
\end{definition}

Intuitively, this definition says that for a functor sending $\C$ to
$\D$ one can choose any morphism $d$ with codomain in the range of $G$
and factor the weakening functor $\C^{!B}$ through the given functor
in such a way that $x$ is sent to $d$.

{\bf Example.\ }  Recall that each object of $\O$ represents a type in the object
programming language.  If we pick some type $T$, then $\O[x{:}T]$ will
be a new category, with an object for every type of $\O$.  The objects
of this new category represent expressions in our object language
having a free variable $x$ of type $T$.  So, for example, if {\tt Int}
is a type, then $\O[x{:}\text{\tt Int}]$ will be the category of
expressions with a free variable $x$ of type {\tt Int}, and if {\tt
  String} is another type, there will be an object
$\O^{!\ttt{Int}}(\ttt{String})$ corresponding to {\tt String} in
$\O[x{:}\text{\tt Int}]$ representing object language expressions
having overall type {\tt String} and a free variable $x$ of type {\tt
  Int}.

If we pick some function $f$ in our object language, where $f$ is a
function that takes an {\tt Int} and returns a {\tt String}, there
will be some $f:\ttt{Int}\to\ttt{String}$ in $\O$.  Now recall that
polynomial categories are just a particular kind of indexed category,
and indexed categories must assign a functor to each morphism
(Definition~\ref{indexed-functor}).  The polynomial category assigns
$f$ a functor $\O^f:\O[x{:}\ttt{String}]\to\O[x{:}\ttt{Int}]$.  Note
that the order of the argument and return type has changed!  This
functor takes a term with a free variable $x$ of type {\tt String} and
yields a term with a free variable $x$ of type {\tt Int}.  How does it
do this?  {\it By substituting $f(x)$ for $x$}.

\subsection{Contextual Completeness}

\begin{figure}
$$
\trfrac
{b:1{\to}B}
{\lift_A(b):A{\to}A{\otimes}B}
\hspace{0.75cm}
\trfrac
{f^B:A\to C}
{\kappa x{:}B. f^B:A{\otimes}B{\to}C}
$$
$$
(\kappa x{:}B. f^B)\circ\lift_A(b) = \subst{x}{b}(f)
$$
\caption{Rules of the $\kappa$-calculus, from \cite{Hasegawa:DecomposingTypedLambda}\label{kappa-calculus}}
\end{figure}

\begin{definition}[{\cite{Lambek:FunctionalCompleteness}}]
\label{contextually-complete}
A polynomial category is said to be {\it contextually complete} if its
weakening functors each have a left adjoint.
\end{definition}

The left adjoint functor will be written $(-){\otimes}B \dashv \C^{!B}$.
The unit of the adjunction $\eta_{-{\otimes}B}:(-){\to}(-){\otimes}B$
has the property that for every $f^B{:}A{\to}C$ in $\C[x{:}B]$ there exists
a $\hat f{:}A{\otimes}B{\to}C$ in $\C$ such that $f^B=\C^{!B}(\hat
f)\circ\eta_{A{\otimes}B}$.  Writing $\lambda x{:}B.f^B$ for
$\hat f$ gives:
$$
f^B = \C^{!B}(\lambda x{:}B.f^B) \circ\eta_{A{\otimes}B}
$$

\begin{remark}
In \cite{Lambek:FunctionalCompleteness}, an explicit definition of
$\lambda f^B$ is given for any contextually complete category {\it
  which also has finite products}; the definition assumes the monoidal
structure of $\C$ has projection and morphism-tupling.  The
construction bears much similarity to typed combinator conversion, but
-- as that author notes -- is completely first-order (in contrast to
Curry's \cite{CurryFeys:CombinatoryLogic} combinator conversion) and
avoids introducing divergent terms (in contrast to Sch\"oenfinkels
\cite{Schoenfinkel}).
\end{remark}

\def\subst#1#2{{[#1{\ttt{:=}}#2]}^{\tsf{Id}}}
Now, select some morphism $b{:}1{\to}B$ and generate the functor
$\subst{x}{b}(-)$ by Definition~\ref{def-poly} corresponding to
the identity functor on $\C$.  It has the following property:
\begin{align*}
             f^B  &= \C^{!B}(\lambda x{f}:B.f^B) \circ\eta_{A{\otimes}B} \\
\subst{x}{b}(f^B) &= \subst{x}{b}(\C^{!B}(\lambda x{:}B.f^B) \circ\eta_{A{\otimes}B})\\
\subst{x}{b}(f^B) &= \subst{x}{b}(\C^{!B}(\lambda x{:}B.f^B)) \circ \subst{x}{b}(\eta_{A{\otimes}B})\\
\subst{x}{b}(f^B) &= ((\subst{x}{b}{\circ}\C^{!B})(\lambda x{:}B.f^B)) \circ \subst{x}{b}(\eta_{A{\otimes}B})\\
\subst{x}{b}(f^B) &= \tsf{Id}_{\C}(\lambda x{:}B.f^B) \circ \subst{x}{b}(\eta_{A{\otimes}B})\\
\subst{x}{b}(f^B) &= (\lambda x{:}B.f^B) \circ \subst{x}{b}(\eta_{A{\otimes}B})
\end{align*}
The last two steps exploit the universal property
$\subst{x}{b}{\circ}\C^{!B}=\tsf{Id}_\C$ of the weakening functor
(Definition~\ref{def-poly}).

Following \cite{Hasegawa:DecomposingTypedLambda}, abbreviate
$\lift_A(b)\defeq\subst{x}{b}(\eta_{A{\otimes}B})$.  The above
definitions and derivations give the three rules of the
$\kappa$-calculus introduced in \cite{Hasegawa:DecomposingTypedLambda}
to isolate the ``first order'' element of the lambda calculus.  These
rules are shown in Figure~\ref{kappa-calculus}.

These inference rules define the syntax of the $\kappa$-calculus, and
the derivation shows that any syntactical term of the calculus
identifies a morphism in a contextually complete category.
The $\kappa$-calculus is a syntax for the internal language of a
contextually complete category in the same way that $\lambda$-calculus
is a syntax for the internal language of a cartesian closed category.

\subsection{Reification}

Having reviewed polynomial categories and the standard definition of
contextual completeness, how can one reason about programs which {\it
  manipulate} other programs with free variables?  Answer: {\it
  reification} of categories.

Just as polynomial categories were a particular kind of indexed
category, reification of one category in another is a particular kind
of {\it indexed functor} between their polynomial categories.

\def\substx#1#2{{\langle}#1{\ttt{:=}}#2{\rangle}}
\def\ot{{\ttt{**}}}
\def\ra{{\ttt{{\tild}>}}}

\begin{definition}
\label{definition-reification}
If $\O[x{:}-]$ and $\M[x{:}-]$ are polynomial categories and
$\code{\cdot}{:}\O\to\M$ is a functor, $\M$ {\it reifies}
$\O$ via $\code{\cdot}$ if there is an indexed functor
$$
\code{\cdot}^{(-)}:\O[x{:}-]\to\M[x{:}\code{-}]
$$
such that for each object $B$ of $\O$ the following diagram commutes
up to isomorphism of functors:
\begin{center}
\begin{tikzpicture}[descr/.style={fill=white,inner sep=2.5pt}]
   \matrix (m) [matrix of math nodes, row sep=3em, column sep=3em] {
     \O[x{:}B] & \M[x{:}\code{B}] \\
     \O        & \M               \\ };
   \path[->,font=\scriptsize] 
    (m-2-2) edge node[auto,swap] {$\M^{!\code{B}}$}  (m-1-2)
    (m-1-1) edge node[auto]      {$\code{\cdot}^B$} (m-1-2)
    (m-2-1) edge node[auto,swap] {$\code{\cdot}$} (m-2-2)
    (m-2-1) edge node[auto]      {$\O^{!B}$}  (m-1-1)
   ;
\end{tikzpicture}
\end{center}
\end{definition}

\begin{remark}
Two technicalities must be noted, but can be skipped on a first
reading.  First, the above abuses notation somewhat: $\code{\cdot}$ is
not strictly the same thing as $\code{\cdot}^{(-)}$; the former is a
non-indexed functor, the latter an $\O$-indexed functor.  The notation is recycled
because the two have similar effect.  Second, $\M[x{:}-]$ is not the
same thing as $\M[x{:}\code{-}]$; the latter is the indexed category
resulting from {\it reindexing} the former along the functor
$\code{\cdot}$.  Similar notation was chosen in order to
de-emphasize the least important details.
\end{remark}

{\bf Example.\ }  Let $\M$ be a category whose objects are the types of the metalanguage
and whose morphisms are its functions; this means that $\M[x{:}-]$ has
an object for every type of the {\it metalanguage}.  The functor
$\code{\cdot}:\O\to\M$ must assign a {\it metalanguage} type to each
{\it object language} type, so in a certain sense the metalanguage has
a copy of the object language type system within it.  Reindexing the
polynomial category $\M[x{:}-]$ by $\code{\cdot}$ to form
$\M[x{:}\code{-}]$ essentially means focusing attention on the subset
of our metalanguage whose free variable types and return types are all
drawn from this copy of the object language's types.
Now, consider the properties bestowed by the indexed functor.
For any object $B\in\O$, the component of the indexed functor will
give a non-indexed functor
$$
\code{-}^B : \O[x{:}-] \to \M[x{:}\code{-}]
$$
What does this functor do?  The last part of
Definition~\ref{definition-reification} requires that the functor
supplied for each object has essentially the same behavior as the
$\code{\cdot}$ functor combined with $\M[x{:}-]$'s weakening functor
$\M^{!B}$.  So if $X$ is an object of $\O$ and $\O^{!B}(X)$ is the
result of weakening $X$ into $\O[x{:}B]$, then reifying this give the
same thing as weakening $\code{X}$ into $\M[x{:}\code{B}]$:
$$\code{\O^{!B}(X)}^B \cong \M^{!\code{B}}(\code{X})$$
This is why similar notation was chosen for $\code{\cdot}$ and
$\code{\cdot}^{(-)}$.
Definition~\ref{indexed-functor} says that for a morphism
$f{:}X{\to}Y$ in $\O$, there will be a functor
$\O^f:\O[x{:}Y]{\to}\O[x{:}X]$.  It was determined earlier that this
functor has the effect of substituting $f(x)$ for $x$ in a term that
has a free variable $x$.  Moving now to the reification functor, it is
clear that $\code{f}^B:\M[x{:}\code{Y}]{\to}\M[x{:}\code{X}]$.  But
what does {\it this} functor do?

Recall that an indexed functor also assigns a natural isomorphism to
every morphism.  Suppose $B$ is an object in $\O$, and $X$, $Y$
are objects in $\O[x{:}B]$.  Then by Definition~\ref{indexed-functor},
our reification functor must assign to each $f:X\to Y$ a natural
isomorphism
$$
\code{-}^f : (\M^{\code{f}}\circ \code{-}^{Y})\cong(\code{-}^{X}\circ\O^f)
$$
This is the key to understanding what $\code{f}^B$ does.  In prose,
the above isomorphism says that applying $\O^f$ and then reifying is
the same as reifying {\it first} and then applying $\code{f}$.  So we
know that $\code{f}$ has the effect of substituting {\it under the
  brackets}, which is exactly the operation needed in order to
manipulate object-language programs.

To sum up, starting from a given functor $\code{\cdot}:\O\to\M$,
asking for a family of functors, one $\code{\cdot}^B$ for each
$B\in\O$ does not say much: these could all be trivial functors which
send every object to a single object and every morphism to its
identity.  Requiring that this family of functors {\it forms an
  indexed functor} is what forces $\code{\cdot}^{(-)}$ to have the
``substitution under brackets'' behavior.  The natural isomorphism
required by Definition~\ref{indexed-functor} turns into precisely the
condition which characterizes the code-splicing behavior of staging
annotations.

\subsection{Contemplation}

\begin{definition}
\label{contemplation}
A category $\M$ {\it contemplates} a category $\O$ if $\M$ reifies
$\O$ and $\M$ is contextually complete.
A category is {\it contemplatively complete} if it contemplates itself.
\end{definition}

Contemplation is the categorical property which best models
multi-stage type systems; Contemplative completeness is the
categorical property which best models {\it homogeneous} multi-stage
type systems.

\begin{theorem}[Staging and Contemplation]
The category whose objects are the types of Figure~\ref{core} and
whose morphisms are the functions definable in that system forms a
contemplatively complete category.
\begin{proof}
Establish a category $\M$ with an object for each type of the language
and for each object $B$ freely generate the polynomial category over
$\M$ in $B$.  The inference rules $\tsf{Lam}$, $\tsf{App}_0$ and
$\tsf{App}_{n+1}$ define the operations of the $\kappa$-calculus and
satisfy the laws of Figure~\ref{kappa-calculus}, so contextual closure
is straightforward.  The syntactical operation which sends an
expression $e$ having free variable $x$ of type $B$ to the expression
$\code{e[x\ttt{:=}(\ttt{\tild}x)]}$ is an indexed functor (with $B$
being the index) whose action on types sends $\M^{!B}(A)$ to
$\M^{!\code{B}}(\code{A})$.  This indexed functor is the reification
functor with the required properties.
\end{proof}
\end{theorem}

\begin{definition}[\cite{Kelly:EnrichedCategories}]
For a monoidal category $\C$ and endofunctor $F:\C\to\C$,
the endofunctor has {\it functorial strength} if for every pair of
objects $A$, $B$ of $\C$ there is a morphism satisfying certain
coherence conditions:
$$
F_{A,B} : F(A){\otimes}B \to F(A{\otimes}B)
$$
\end{definition}

\begin{definition}
A contemplatively complete category has {\it enriched contemplation}
if the coordinates of the reification functor all have strength.
\end{definition}

Strengths on the reification functor give the ability to perform
cross-stage persistence.  The morphism $\code{\cdot}_{1,A} :
\code{1}{\otimes}A\to \code{1\otimes A} = A\to \code{A}$ provides the
required transition.

\subsection{$\kappa$-Categories and Freyd Categories}

\begin{definition}[{\cite[Definition 11]{PowerThielecke:KappaCategories}}]
A $\kappa$-category consists of a finite product category $\C$ and a
$\C$-indexed category $H^{(-)}$ such that:
\begin{enumerate}
\item For each object $A$ of $\C$, $H^A$ has the same objects as $\C$,
      and $H^f$ is the identity on objects.
\item For each projection morphism $\pi:B{\times}A{\to}B$ of $\C$,
      $H^\pi$ has a left adjoint $(-){\times}A$
\item For each morphism $f:B\to B'$, the natural transformation
      $\phi:((-){\otimes}B)\circ H^{f\times\tsf{id}_A} \to H^f \circ
      ((-){\otimes}B')$ induced by the adjointness in the previous
      bullet point is in fact an isomorphism.
\end{enumerate}
\end{definition}

\begin{theorem}
Categories with enriched contemplation and finite products are in bijective
correspondence with $\kappa$-categories.
\begin{proof}
Given a category $\M$ with enriched contemplation and finite products,
$\M[x{:}-]$ is the requisite $\M$-indexed category, (1) each
$\M[x{:}B]$ has the same objects as $\M$ and the weakening functor
$\M^{!B}$ is identity-on-objects (Definition~\ref{def-poly}), (2)
because $\M$ is contemplative it is contextually complete
(Definition~\ref{contemplation}), so the weakening $\M^\pi$ of any
projection morphism $\pi$ has left adjoint
(Definition~\ref{contextually-complete}), and (3) the natural
isomorphism imposed by the indexed reification functor
(Definition~\ref{definition-reification}) supplies the requisite
$\phi$.
\end{proof}
\end{theorem}

\begin{definition}[{\cite[A.4]{PowerThielecke:ClosedFreydKappa}}]
A {\it Freyd Category} is a category $\C$ with finite products, a symmetric
premonoidal category $\K$, and an identity-on-objects strict symmetric
premonoidal functor $J:\C\to\K$.
\end{definition}

\begin{theorem}[{\cite[Theorems 13 and 14]{PowerThielecke:KappaCategories}}]
Freyd Categories and $\kappa$-categories and are in bijective
correspondence.
\end{theorem}

\begin{theorem}[The Stages-Arrows Isomorphism]
\label{stages-arrows-isomorphism}
Categories with enriched contemplation and finite products are in bijective
correspondence with Freyd categories.
\begin{proof}
By transitivity of bijective correspondence.
\end{proof}
\end{theorem}

\begin{remark}
The proof shown for Theorem~\ref{stages-arrows-isomorphism} is clearly
trivial once the appropriate context has been set up.  The main
contribution of this section is not a one-line proof, but rather the
identification and definition of {\it enriched contemplation} as the
appropriate criterion.  Specifically, enriched contemplation is a
strong enough condition to make the proof of bijective correspondence
go through (almost effortlessly), but still weak enough that a large
class of stage-annotated metaprogramming languages constitute
categories with enriched contemplation.  Furthermore, enriched
contemplation is not even quite so important as the weaker forms it
suggests.  If categories with enriched contemplation and finite
products are in bijective correspondence with Freyd categories, it is
natural to ask what is in bijective correspondence with obvious
weakenings such as monoidal categories with enriched contemplation,
premonoidal categories with enriched contemplation, categories with
non-enriched contemplation, and categories which reify categories
besides themselves.  Generalized arrows subsume all of these.  So
while Theorem~\ref{stages-arrows-isomorphism} may not be surprising or
unlikely, the connection it establishes justifies the generalization.
\end{remark}

\begin{figure*}[t]
{\footnotesize{
\begin{verbatim}
  id_left    : forall (A B:Set)    (f:A~>B),                                id >>> f ~~ f
  id_right   : forall (A B:Set)    (f:A~>B),                                       f ~~ f >>> id
  comp_assoc : forall (A B C D:Set)(f:A~>B)(g:B~>C)(h:C~>D),         (f >>> g) >>> h ~~ f >>> (g >>> h)
  first_law  : forall (A B C D:Set)(f:A~>B)(g:B~>C),                 first (f >>> g) ~~ first(c:=D) f >>> first g
  law5       : forall (A B C:Set)  (f:A~>B),               first (first f) >>> assoc ~~ assoc(c:=C)(b:=B) >>> first f
  law6       : forall (A B C:Set),                                             cossa ~~ swap >>> assoc (b:=B) >>> swap
  law7       : forall (A B C:Set)(f:A~>B),                          first f >>> drop ~~ drop (b:=B) >>> f
  law8       : forall (A B:Set),                          swap (b:=B)(a:=A) >>> swap ~~ id
  law9       : forall (A B:Set),                                       copy >>> swap ~~ copy (a:=A)
  law_assoc  : forall (A B C:Set),                assoc (c:=C)(b:=B)(a:=A) >>> cossa ~~ id
  law_cossa  : forall (A B C:Set),                cossa (c:=C)(b:=B)(a:=A) >>> assoc ~~ id
\end{verbatim}
}}
\caption{{\tt GArrow} laws of Figure~\ref{laws}, rendered as Coq
  propositions to be satisfied by any {\tt Instance} of {\tt
    GArrow}\label{laws-coq}}
\end{figure*}

\section{Future Work}

\subsection{Polymorphism and Inference}

The presentation in this paper did not cover either type polymorphism
or inference; these will be necessary for a production-quality
system.  This will require extending the grammar for types:
\begin{align*}
\alpha       & ::= \text{\ type variables} \\
\tau         & ::= \ldots \ |\ \alpha\ |\ \forall \alpha . \tau
\end{align*}
The $\fo(\tau,\veta)$, $\tsf{reifiable}(\tau,\veta)$, and
$\tsf{recOk}(\tau,\veta)$ judgements present a small complication for
polymorphism; when attempting to assign a polymorphic type to an
expression, the typical rule used \cite{Taha:ClassifierInference} is
something similar to:
$$
\trfrac{
  \begin{trgather}
  \alpha\notin\tsf{FV}(\G_1,\G_2,\tau_2,\veta) \\
  \G_1 \vdash e_1:\tau_1^\veta \\
  \ctree{\G_2}{x:(\forall\alpha.\tau_1)^\veta} \vdash e_2:\tau_2^\veta \\
  \end{trgather}
 }{
  \ctree{\G_1}{\G_2} \vdash \ttt{let\ }x\ttt{=}e_1\ttt{\ in\ }e_2 : \tau_2^\veta  
 }
$$
In this arrangement, the type inference procedure may find itself
confronted with the need to prove judgements such as
$\fo(\alpha,\veta)$ where $\alpha$ is a type {\it variable}.  The
solution to this situation is to introduce qualified types
\cite{Jones:QualifiedTypes}, gathering a list of constraints imposed
on each type variable and annotating type quantifiers with these
constraints, creating types such as $\forall\alpha . \fo(\alpha,\veta)
\Rightarrow \tau$.

Level polymorphism will also be necessary for a production-quality
system.  The algorithm described in \cite{Taha:ClassifierInference}
appears to be the most appropriate.  Among the changes required will
be extending the grammar for types:
$$
\tau ::= \ldots \ |\ \forall {\eta} . \tau
$$
and adding a typing rule to propagate the $\fo(\tau,\veta)$ judgement
across level quantifiers:
$$
\trfrac[$\text{FC}_\forall$]
  {\begin{trgather}
   \eta'\notin\tsf{FV}(\tau,\veta) \\
   \fo(\tau[\eta\ttt{:=}\eta'],\veta)
   \end{trgather}
  }
  {\fo(\forall\eta.\tau,\veta)}
$$

\subsection{Dependent Types}

The characterization of staging annotations as an indexed functor
among polynomial categories gives a category-theoretic foundation to
multi-stage programming.  In this context, dependent types are
understood as the objects of locally cartesian closed categories
\cite[Definition 9.19]{Awodey:CategoryTheory}.  This should provide a
straightforward way to investigate multi-stage programming at all
corners of the lambda-cube \cite{Barendregt:LambdaCube}, perhaps
leading to a sound multi-stage Calculus of Constructions \cite{CoquandHuet:CoC}.

\bibliographystyle{alpha}
\bibliography{megacz-icfp10}


\end{document}